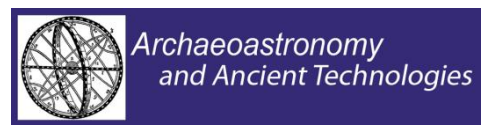

# The Prototype of Ancient Analemmatic Sundials (Rostov Oblast, Russia)

**Larisa N. Vodolazhskaya[1], Pavel A. Larenok[2], Mikhail Yu. Nevsky[3]**

[1] Southern Federal University (SFU), Rostov-on-Don, Russian Federation;
E-mails: larisavodol@aaatec.org, larisavodol@gmail.com
[2] NP "Yuzharheologiya", Rostov-on-Don, Russian Federation; E-mail: dao2@inbox.ru
[3] Southern Federal University (SFU), Rostov-on-Don, Russian Federation; E-mails: munevsky@sfedu.ru

**Abstract**

The article presents the results of a study of petroglyphs on a unique stone slab discovered near the kurgan 1 of the kurgan field Varvarinsky I (Rostov Oblast, Russia). During the study was done comparing it with plates of Srubna burials on which were depicted the petroglyphs. Similar features for all the considered slabs are elliptically arranged wells. Groove made to Varvarinsky stone slab, has the same parameters as the ellipse from the wells on the slab of the kurgan field Tavriya -1, discovered in the Rostov region, too. With the help of astronomical methods previously been proved that elliptically located wells on the Srubna slabs are hour markers of analemmatic sundial. By results of the use of similar methods to study complex of petroglyphs on Varvarinsky slab, it was concluded that the slab could be used for development of the technology of markup of sundial on different geographical latitudes. Also, Varvarinsky slab could serve as a model example for a markup of the ellipse on which subsequently extorted wells for latitudes, close to the latitude of the place of slab detection. Thus, Varvarinsky stone slab is the prototype of analemmatic sundials, discovered in the Rostov and Donetsk oblasts.

## Introduction

In 2013, archaeologist Anatoly Fayfert discovered the stone slab with petroglyphs in Sholokhov district of Rostov Oblast (Russia) (Fig. 1). The slab was found in 30 m to the southeast of kurgan 1 of the kurgan field Varvarinsky I (49.50 N, 41.40 E). Kurgan has a height of 0.4 m at the moment, and it is pulled on the North-South line. Plate weight around 70 kg. The second side of slab does not contain petrogliphs [1, p. 27-28]. Author of find suggests that the slab could be related to the burial rite or to the sanctuary. Slab consists of coarse-grained sandstone red-brown color.

Approximately 300 meters to the north of kurgan 1 is a natural deposit of quartzite slabs. Further north, near the hamlet Varvarinsky, a large deposit of quartzite exists and has name is "Stone Mountain". Quartzites and sandstones are typical for a large part between the rivers Don and the Seversky Donets. To the north, close to the Don, quartzite occurs in small patches in the form of compact clusters of different sizes: the wrong discs and slabs, the gaps between them are filled with loose arenaceous sediments and covered with sod [2, p. 299].

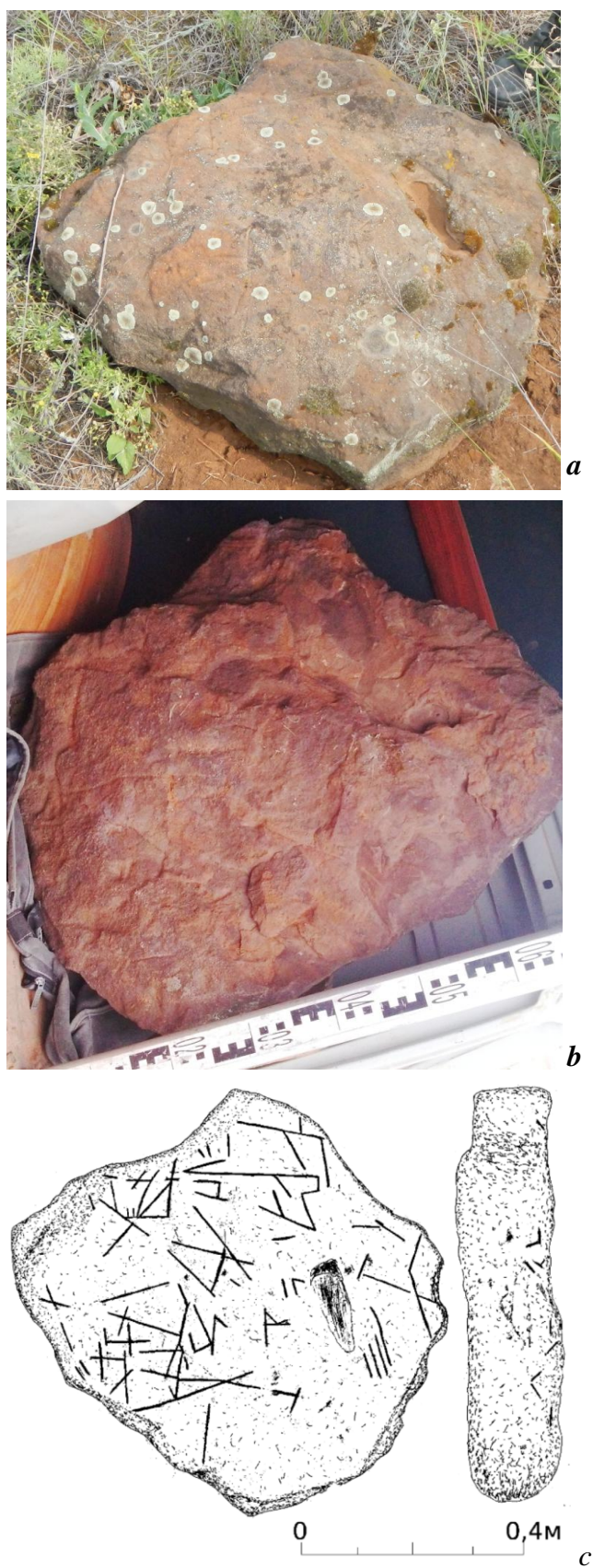

**Figure 1.** Kurgan field Varvarinsky I, kurgan 1 (neighborhood), stone slab with petroglyphs: *a* - the upper side of the slab (photo by A.V. Fayfert, 2013) [1, Fig. 21], *b* - the bottom side of the plate (photo by A.V. Fayfert, 2013), *c* - drawing the top slab side (drawing by A.V. Fayfert, 2015) [1, Fig. 22].

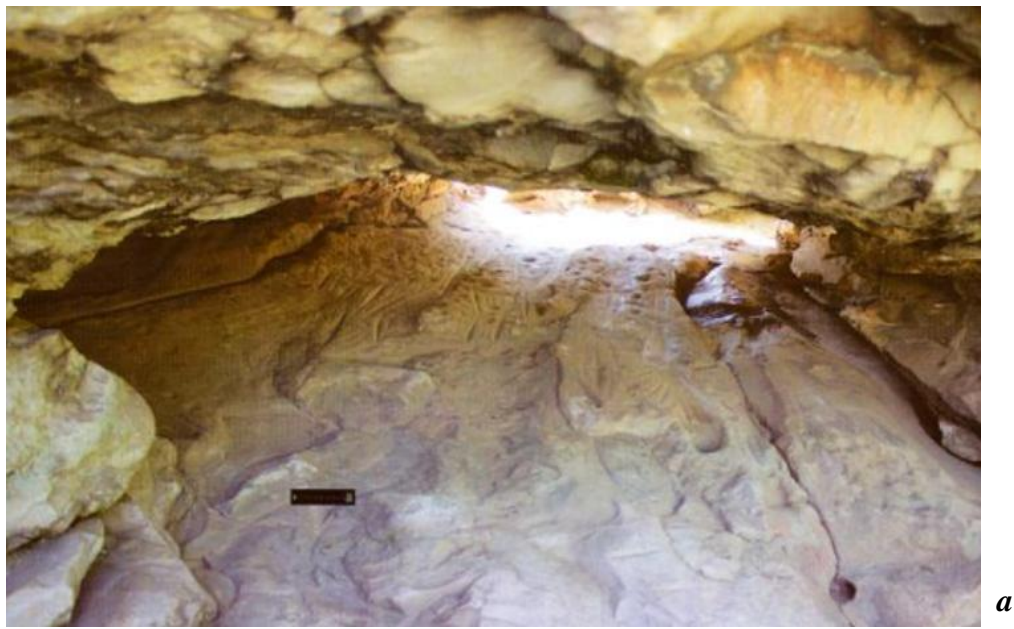

*a*

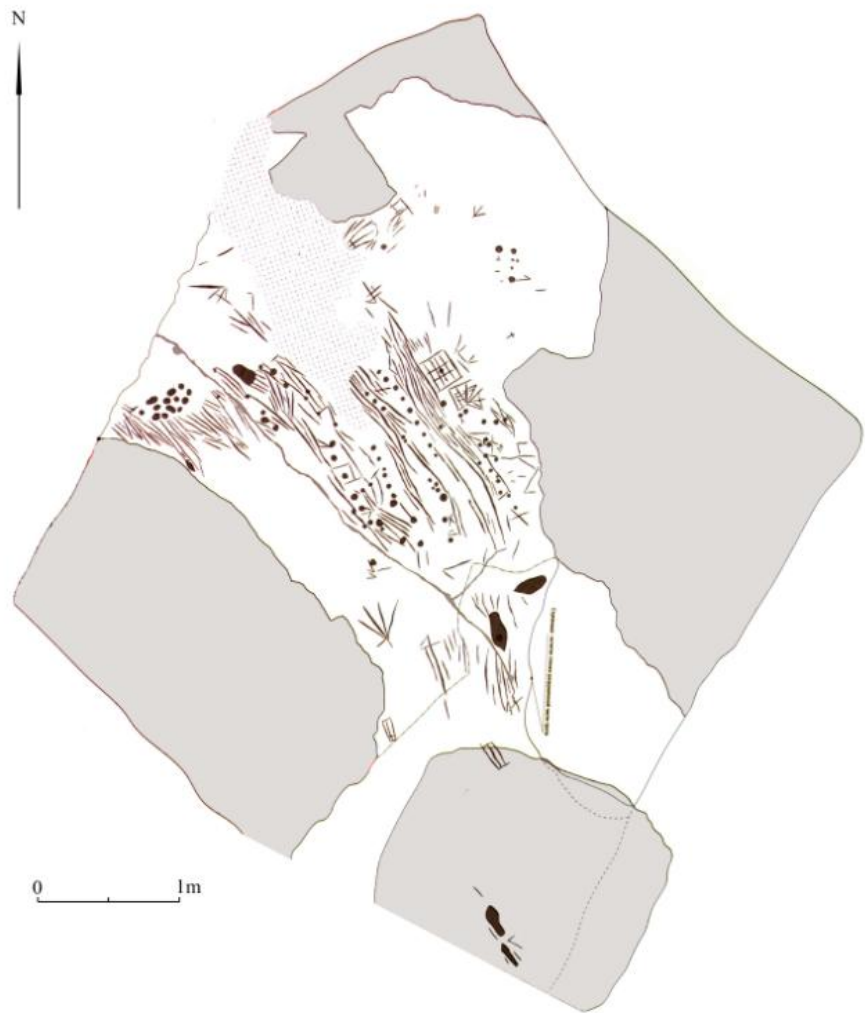

*b*

**Figure 2.** Skelnovsky grotto with petroglyphs: *a* - photo of the grotto [3, Fig. 14], *b* - portrayal of grotto petroglyphs [3, Fig. 12].

In the north of Rostov region, in the vicinity of hamlet Skelnovsky, in a massive quartzite block was discovered grotto with petroglyphs (Fig. 2). It was dated from the end of IV millennium BC before the usual kurgan cultures of developed Bronze Age [3, p. 17]. Because Varvarinsky hamlet is not far from hamlet Skelnovsky (less than 30 km), Varvarinskaya slab was dated by the author of find by analogy with skelnovskimi petroglyphs [1, p. 44-45]. Skelnovsky grotto is the object of the cult practices of pastoralists of the Early Metall age [3, p. 16] and, possibly, a kind of stone chronicle of bright astronomical phenomenon - the fall of a large meteorite similar to Sikhote-Alin meteorite, accompanied by a meteor shower [4].

**Srubnaya slabs with petroglyphs**

To date, three slabs of fine-grained sandstone with petroglyphs were found in two adjacent areas - Rostov and Donetsk Oblast. They belong to the Srubna culture and date back to the period of the Late Bronze Age (approximately, XVII- XII cent. BC).

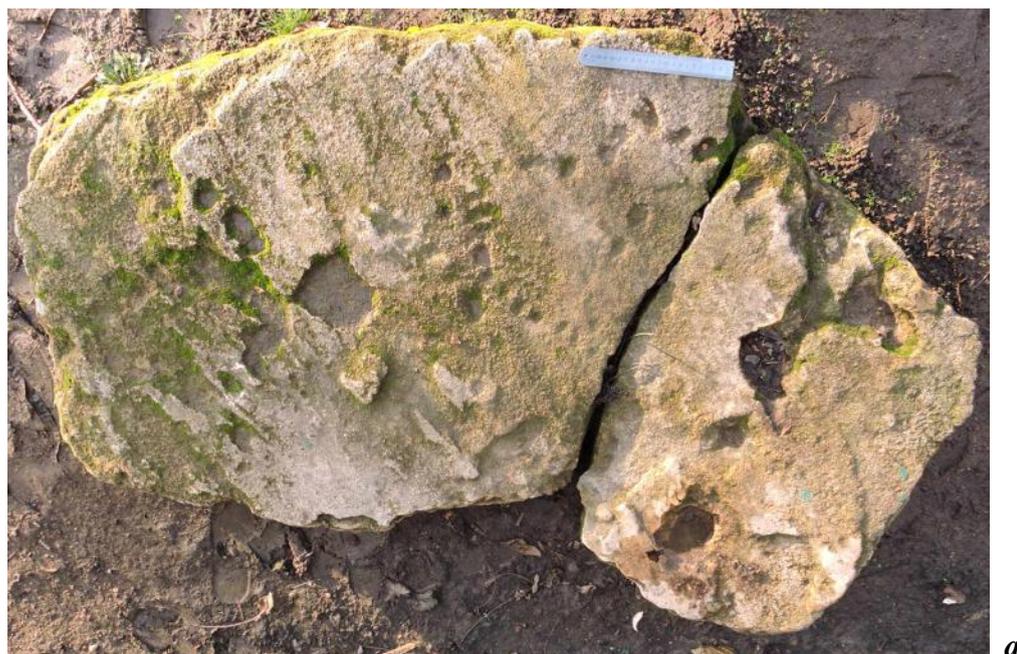

*a*

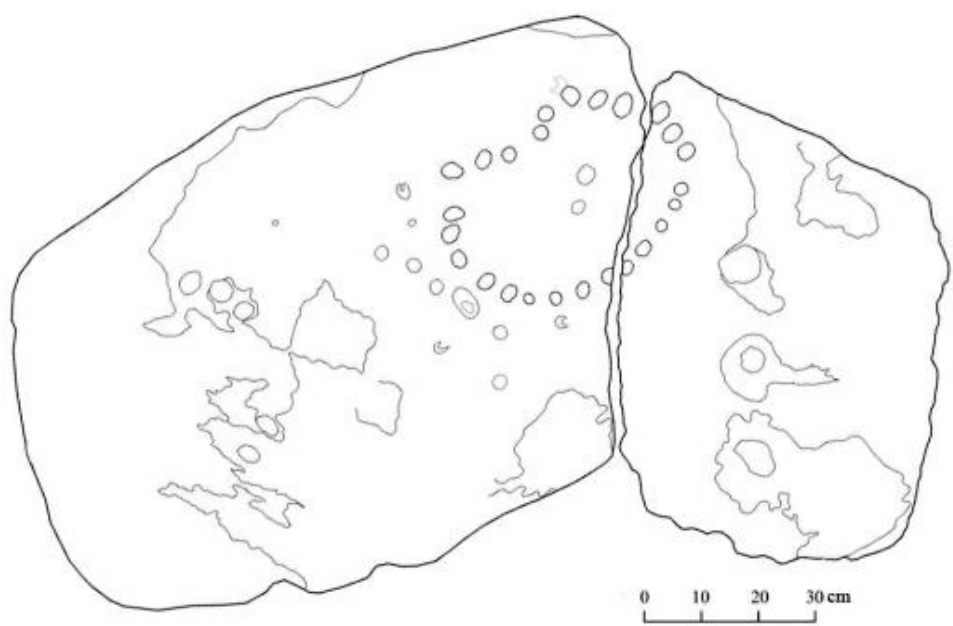

*b*

**Figure 3.** Kurgan field Tavriya-1, kurgan 1, burial 2, floor slab with petroglyphs: *a* - photography of the slab (photo by A.I. Mishchenko, 2016), *b* - drawing of the slab [6, fig. 4].

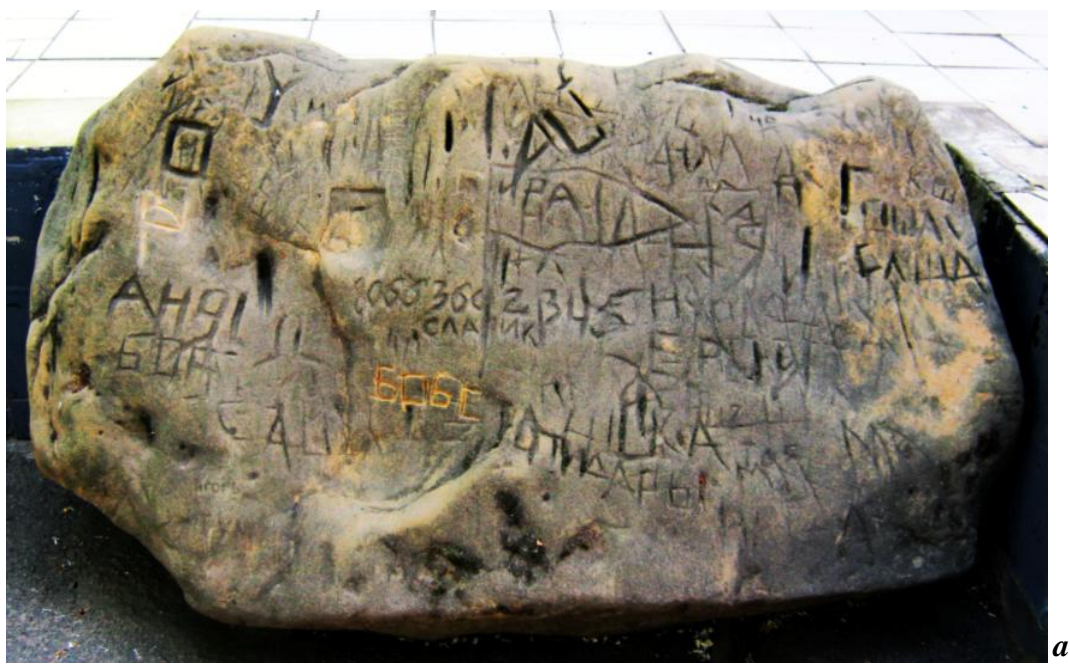

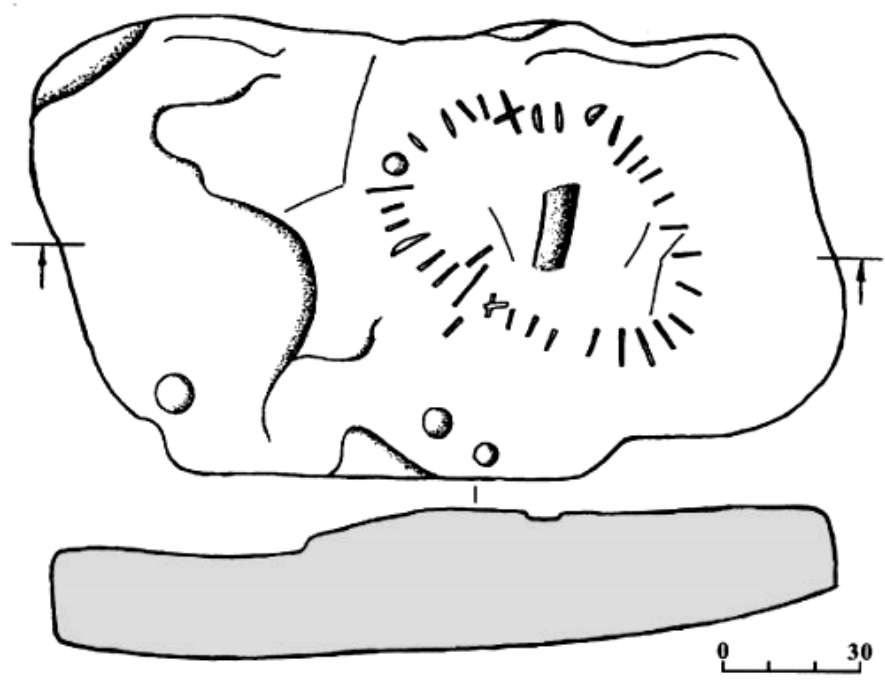

**Figure 4.** Kurgan group Rusin Yar, kurgan 1, burial 1, slab of burial construction with petroglyphs: *a* - photo of the slab (photo by Yu.B. Polidovich, 2011), *b* - drawing of the slab [10, Fig. 2].

One slab - from the kurgan field Tavria-1 of the Rostov Oblast[3] (Fig. 3) [5, p. 62], [6] and the two slabs - from a kurgan groups Rusin Yar[4] (Fig. 4) and Popov Yar-2[5] (Fig. 5) of the Donetsk Oblast [7], [8, p. 444-455], [9, c. 36-135].

---

[3] The slab is stored in the territory of the Archaeological Museum - reserve "Tanais" (Rostov Oblast., Russia)
[4] The slab is on the unprotected territory near the building of the Konstantinovsk City Museum (Donetsk Oblast, Ukraine).
[5] The slab is stored in Donetsk regional museum (Donetsk Oblast, Ukraine).

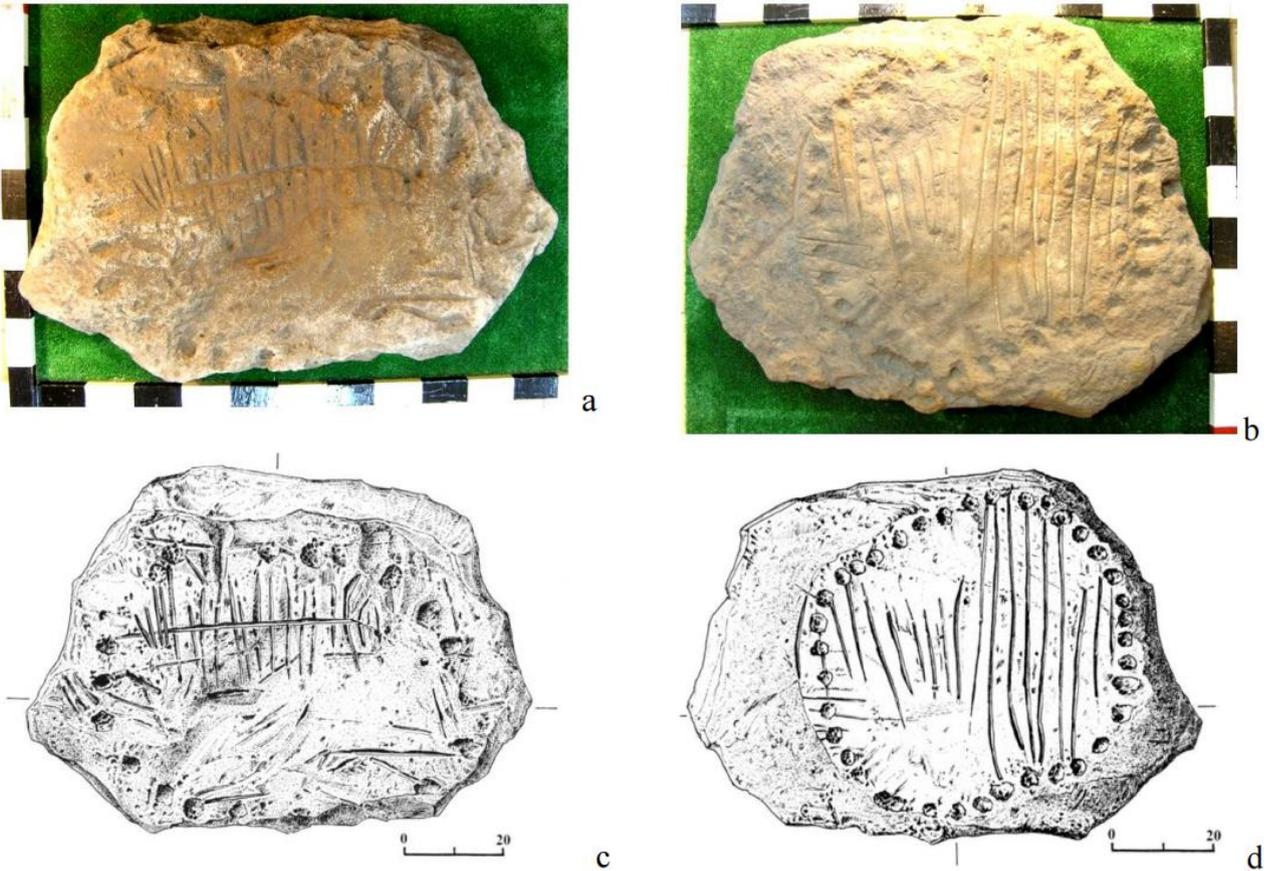

**Figure 5.** Kurgan group Popov Yar-2, kurgan 3, burial 7, floor slab with petroglyphs: *a* - photo of side A [9, Fig. 44.1], *b* - photo of side B [9, Fig. 45.1], *c* - drawing of side A [9, Fig. 42.1], *d* - drawing of side B [9, Fig. 42.2].

**Comparative analysis of petroglyphs**

An analysis of the petroglyphs on slabs from Popov Yar-2 and Tavria-1 (Fig. 6) was carried out with the help complex methods of natural science, which already shown to be effective in the interdisciplinary research of ancient structures and artifacts [11-20]. As a result, it was proved that the petroglyphs in the form of arranged elliptically wells on these slabs are hour markers of analemmtic sundial [6], [21]. Ellipse from the wells located at the slab from Rusin Yar too, but its analysis, in terms of hour marks a sundial, with the required accuracy is not possible because slab drawing was made approx and slab surface is now badly damaged (Fig. 4a).

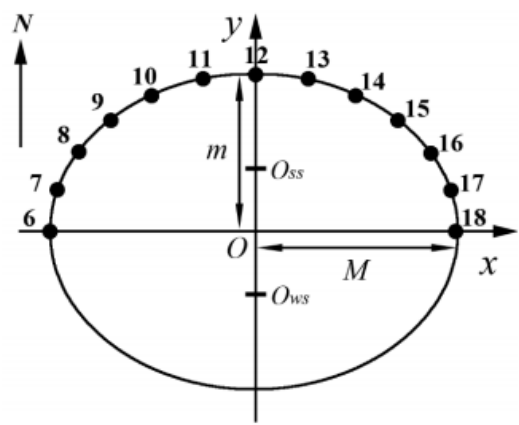

*a*

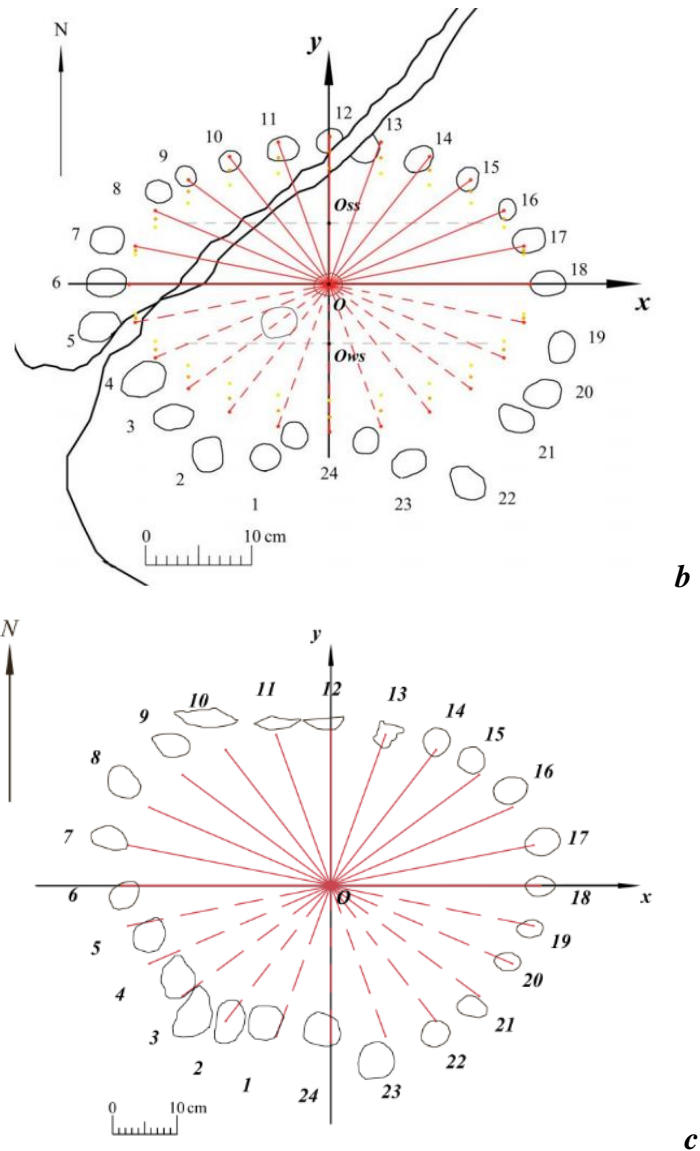

**Figure 6.** Hour markers of analemmatic sundial: *a* - on the coordinate plane [6, fig. 6], *b* – on the floor slab from burial 2, kurgan 1, kurgan field Tavriya-1; hour lines of analemmatic sundial of latitude 47°17' *N* of Tavriya-1 for equinox (red color). For comparison hour markers for latitude 41°00' *N* are marked in orange color, and for the latitude 35°00' N – yellow color [6, рис. 7]; *c* – on the slab from burial 7, kurgan 3, kurgan group Popov Yar-2; hour lines for Equinox for latitude of Popov Yar-2, equal to 48°26' *N*, indicated by the red lines. Dashed lines - hour lines of nonworking range. *N* - True North.

Slab from Popov Yar-2 dimensions are 60×90÷95×9÷23 cm, and the ellipse from the wells on it is 62÷70×45÷50 cm, slab from Rusin Yar dimensions are 85÷98×158÷162×20÷30 cm approximate and the ellipse from the wells is 60÷70×40÷50 cm approximately, slab from Tavriya-1 dimensions are of 85×164×15÷20 cm, and undamaged ellipse from the wells on it is 38÷44×25÷29 cm. On Varvarinsky slab engraved wells are absent, but in the center of the slab a resembling an ellipse groove is easily visible. It is made approximately by three quarters of the ellipse length (Fig. 7). The width of the groove at an average of about 4 cm, depth is not uniform and reaches ≈1 cm. In the bottom left of the slab groove is not visible. Varvarinsky slab approximate dimensions are 60×75×10÷15 cm, and the ellipse approximately is 40×30 cm.

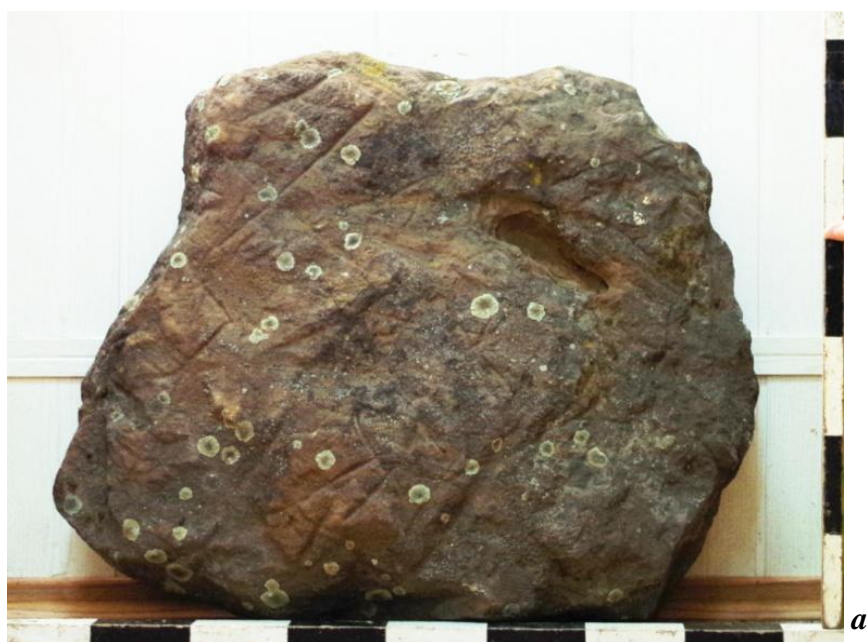

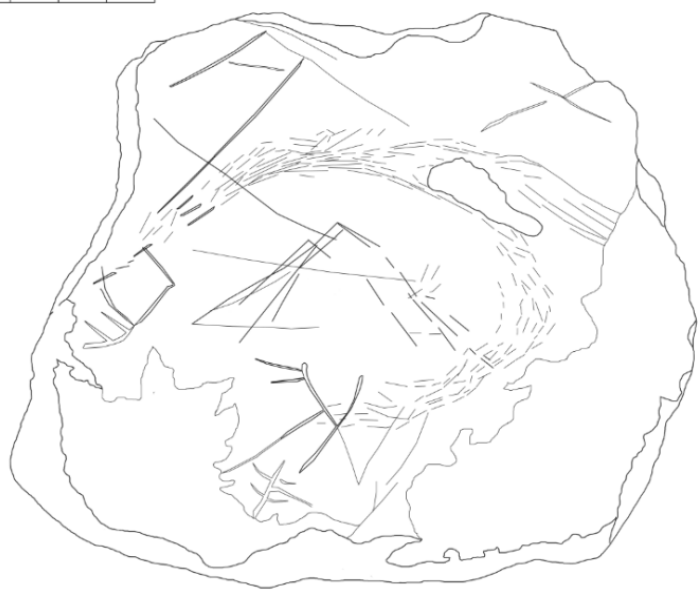

**Figure 7.** Kurgan field Varvarinsky I, kurgan 1 (neighborhood), a stone slab with petroglyphs: *a* - photo of the slab[6] (by A.D. Vodolazhskiy, 2014), *b* – drawing of the slab (by L.N. Vodolazhskaya, 2015). Slab is worth perpendicular to the floor (the alignment was carried out using plumb lines, slab supported by special fasteners located between the slab and the wall).

Analysis of the dimensions of these slabs with petroglyphs shows that the dimensions of slab from Rusin Yar similar to dimensions of slab from Tavria-1, and the dimensions of Varvarinsky slab dimensions similar to dimensions of slab from Popov Yar-2. Moreover, these two slabs consist of ferruginous sandstone. The ellipse from the wells dimensions on the slab from Rusin Yar approximately match the dimensions of the ellipse on the slab from Popov Yar-2 and approximately in one and a half times the dimensions of the intact ellipse on the slab from Tavria-1. Dimensions of the the ellipse on the slab from Tavria-1 close to the dimensions the ellipse on the Varvarinsky slab. This is especially noticeable at overlapping contours of the wells on the slab from Tavria-1 with Varvarinsky slab with preservation of the scale (Fig. 8). Wells coincide with a wide groove and the cusps from the wells ("cat ears") are repeated angular protrusions of the upper edge of the Varvarinsky slab.

---

[6] The photo was taken with the camera, Pentax K-50. For reduction of distortions the process of photographing of vertically standing (checked by means of a plumb) slab produced at a distance about four meters. The lens of the camera, directed perpendicular to slab surface, was opposite to its center. The resulting slab image occupied a small part in the center of the frame. Location photometers monitored by building levels. The flash is not used.

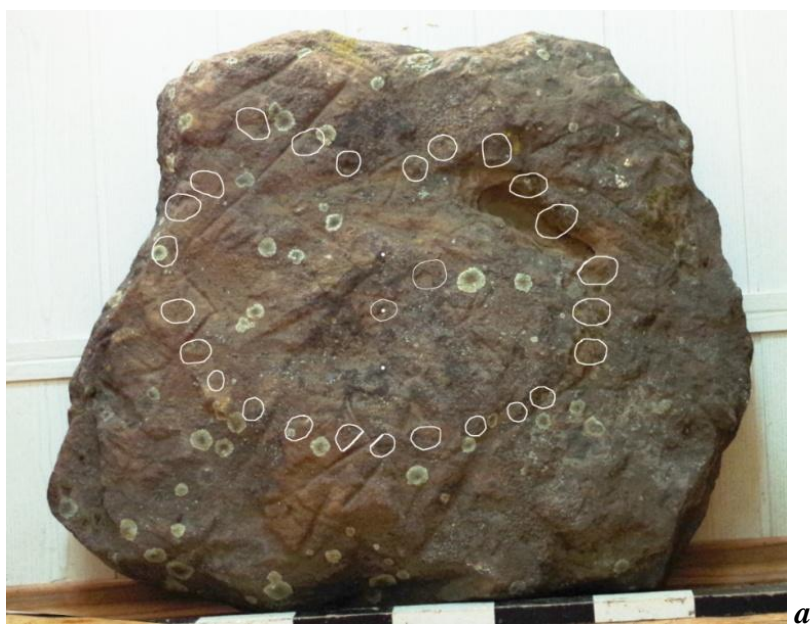

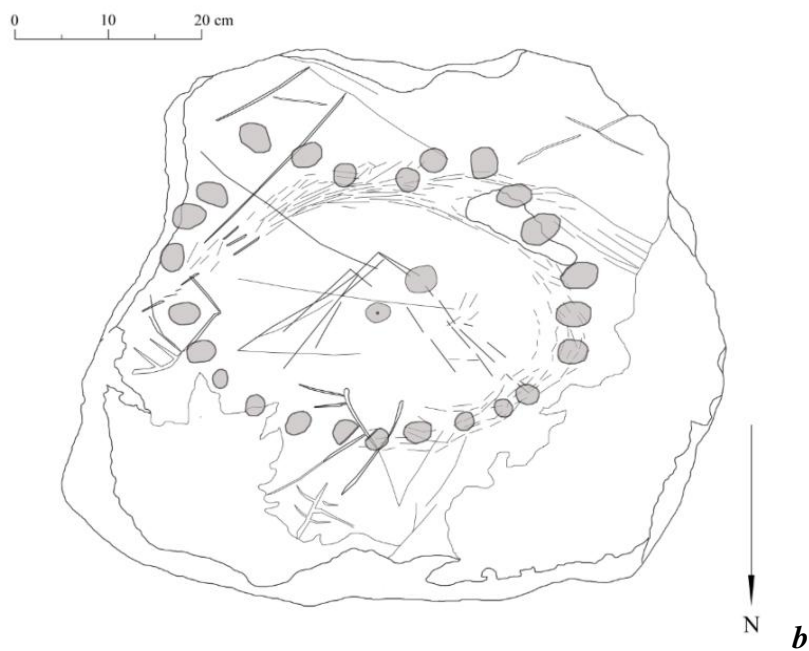

**Figure 8.** Kurgan field Varvarinsky I, kurgan 1 (neighborhood), a stone slab with petroglyphs: *a* - wells of slab from Tavria-1 contours combined with the photo of Varvarinsky slab, *b* - wells of slab from Tavria-1 contours combined with the drawing of Varvarinsky slab.

Protrusions are likely determined an axis of symmetry when marking the slab. Taking as a basis axis of symmetry of the ellipse from the wells on the slab from Tavria-1, we reconstructed the original shape of Varvarinsky slab. As starting profile for a reconstruction of the lost fragment was used the profile of the opposite side of slab. Symmetric reflection profile of the right half of slab (Fig. 9a) not bad match with the profile of the undamaged parts of the left half of the slab. Restored with the help the described method the profile of the lost fragment of slab fits well with the profile of slab, extending it to close to the oval shape (Fig. 9b).

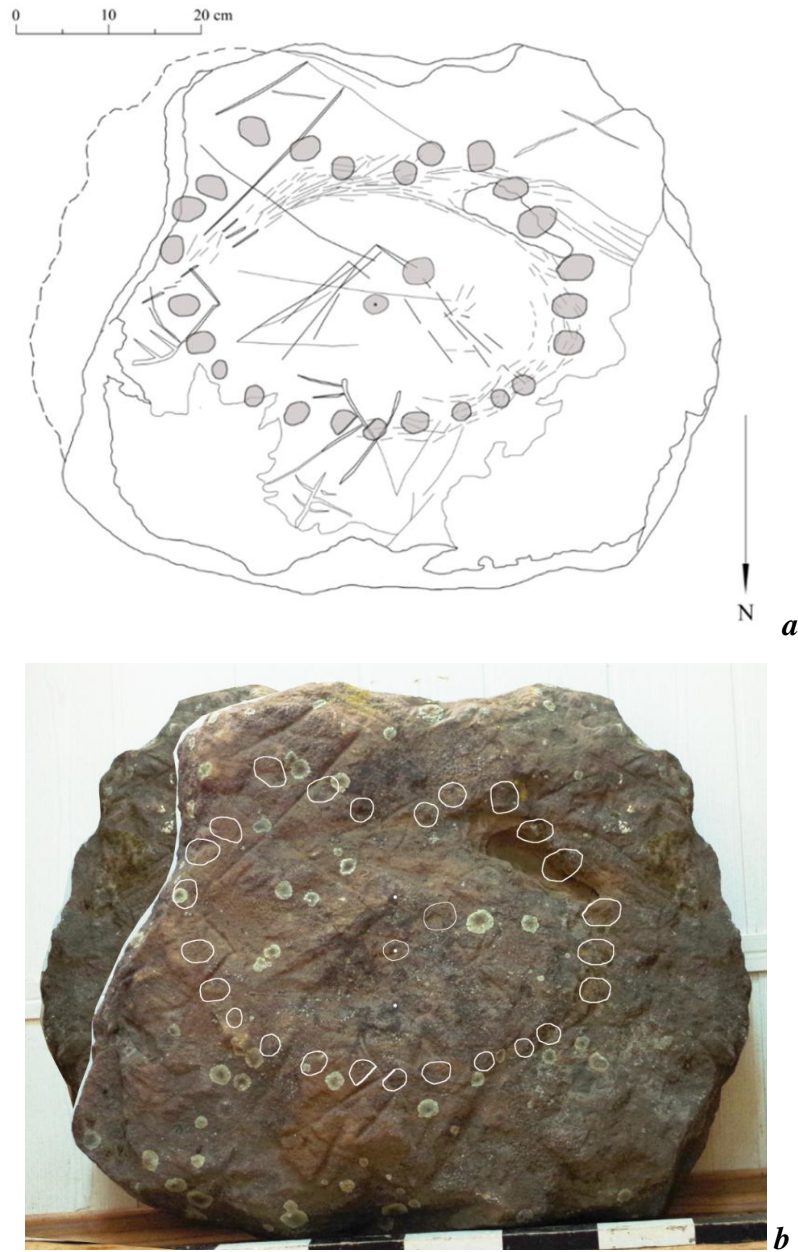

**Figure 9.** Kurgan field Varvarinsky I, kurgan 1 (neighborhood), a stone slab with petroglyphs: *a* – drawing of surface of the slab with the reconstructed profile (dashed line); vertical line in the center of the drawing - an axis of symmetry of the ellipse from the wells, *b* - photo plates with reconstructed of lost part.

The similarity between the dimensions of the ellipse from the wells on the slab from Tavriya-1 and the groove in the form the ellipse on Varvarinsky slab, on the one hand, and the similarity between the dimensions and shape of the sharp projections on the ellipse from the wells and the corresponding angular protrusions of upper edge of Varvarinsky slab, on the other hand, gives grounds to assume that Varvarinsky slab as well as a slab from Tavriya-1 belongs to the Srubna culture, and the groove in the form of the ellipse is directly related to technology of marking of analemmatic sundial.

It should also be noted that a large oblong alveolus of natural origin, likely trace by fossil, located directly on the groove of Varvarinsky slab. Form of alveolus resembles a trace of human foot, so in ancient times this slab can be revered as a sacred stone-'sledovik' or Footprint Stone [22], [23] and therefore this slab could be selected for the application of the petroglyphs.

When combining the ellipse from the wells of slab from Tavriya-1 with a photo and drawing of Varvarinsky slab, can be seen that the well of 18 hours, by which can be measured semi-major axis of the ellipse *M*, adjudged to be in center of the sign "rhombus" (Fig. 10 a, 10 b) and well of 12 hours, by which to measure the semi-minor axis of the ellipse *m*, is projected near the center of the sign "oblique cross" (Fig. 10 c, 10 d).

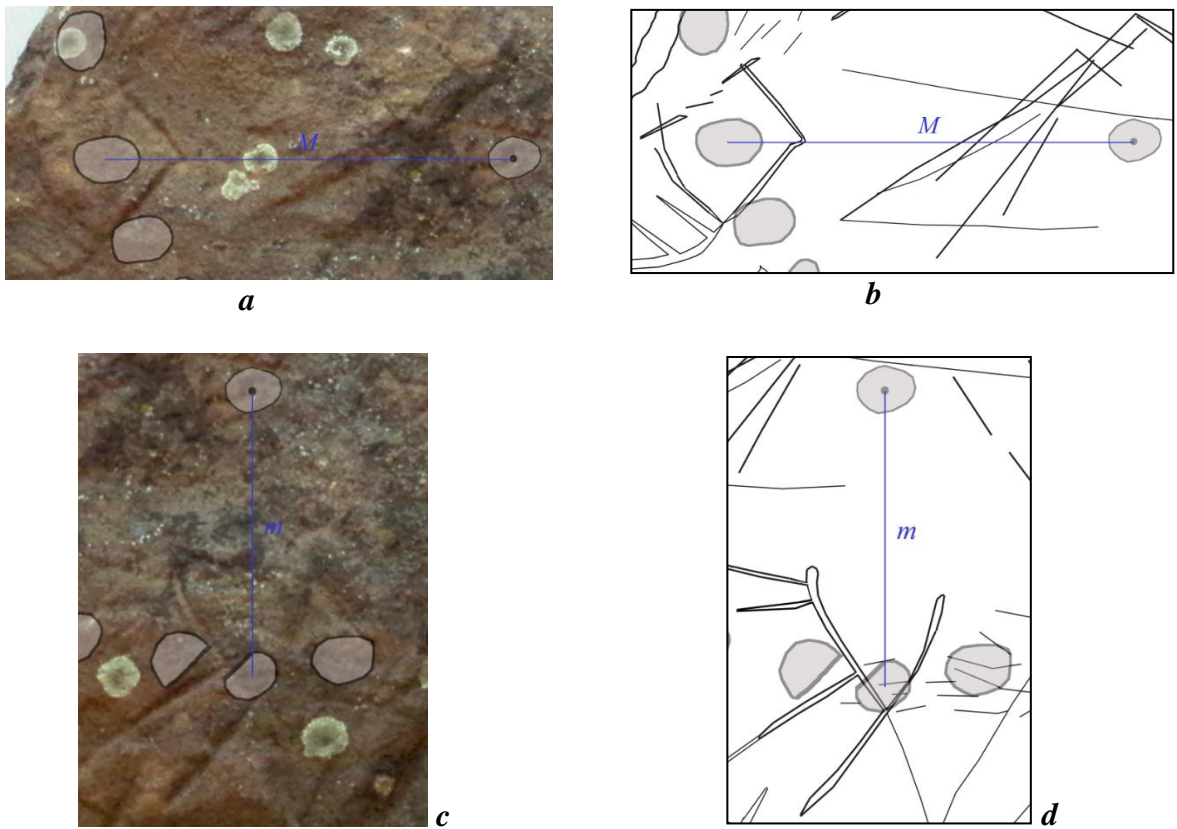

**Figure 10.** Kurgan field Varvarinsky I, kurgan 1 (neighborhood), a stone slab with petroglyphs: *a* – photo of the fragment of slab with petroglyph "rhombus" and semi-major axis of the ellipse of "dial" *M*, *b* - drawing of slab fragment with semi-major axis *M*, *c* - photo of slab fragment with petroglyph "oblique cross" and small semi-axis *m*, *d* - drawing of slab fragment with minor axis *m*.

Line segment by center of the ellipse to the inner edge of wells was seen as semi-major axis, originally in research of slab of Tavriya-1. In another case, the line segment corresponding to the minor axis, extending beyond well of 12 hours. Based on the assumption that the plate is likely to have been made or marked near the site of its discovery, we concluded that the slab layout was made is not very accurate. However, the fact that one of the key well of slab from the Tavriya-1, which determines the semi-major axis of the ellipse *M*, got into the center of sign "rhombus" has forced us to come to the conclusion that the measure and the marking of analemmatic sundial ellipse were based on the distance to the center of the wells.

Measured semi-major axis of the ellipse from the wells on the slab from Tavriya-1 is $M_{Tav} \approx 20.5$ cm (relative center of well). For latitude of Tavriya-1 *Lat*=47°17' N, minor axis of the ellipse of analemmatic sundial calculated according to the formula 1 $m_{Tav}$=15.1 cm (relative center of well).

$$m = M \cdot \sin\varphi, \tag{1}$$

$$x = M \cdot \sin H, \tag{2}$$

$$y = M \cdot \sin\varphi \cdot \cos H, \tag{3}$$

$$Z_{ws} = M \cdot tg\,\delta_{ws} \cdot \cos\varphi, \tag{4}$$

$$Z_{ss} = M \cdot tg\,\delta_{ss} \cdot \cos\varphi, \tag{5}$$

$$H' = arctg(tgH/\sin\varphi), \text{ for } t \in [6; 18] \tag{6}$$

$$H' = arctg(tgH/\sin\varphi) - 180°, \text{ for } t \in [0; 6[$$

$$H' = arctg(tgH/\sin\varphi) + 180°, \text{ for } t \in ]18; 24],$$

$$\text{where } H = 15° \cdot (t - 12),$$

where *x* - coordinate of a point on the *X* axis for analemmatic sundial, *y* - coordinate of the point on the *Y* axis for analemmatic sundial, $M \approx 19$ cm - measured semi-major axis of the ellipse, $\varphi$ - latitude of location, *t* - the true local solar time, *H* - hour angle of the Sun, *H'* - angle between the meridian line and the hour line on the sundial, $\delta_{ws} = -\varepsilon$ - declination of the Sun at the winter solstice, $\delta_{ss} = \varepsilon$ - declination of the Sun at the summer solstice, $y = Z_{ws}$ – in the winter solstice, $y = Z_{ss}$ – in the summer solstice (Fig. 6a).

During the summer solstice, the sun declination equals obliquity of the ecliptic $\delta_{ss}=\varepsilon$, which is calculated using Formuls [24, c. 35]:

$$\varepsilon = 23.43929111° - 46.8150'' \cdot T - 0.00059'' \cdot T^2 + 0.001813 \cdot T^3, \tag{7}$$

$$T \approx \frac{(y - 2000)}{100}, \tag{8}$$

where T - the number of Julian centuries separating the era from noon January 1, 2000, y – the year. During the winter solstice the sun's declination $\delta_{ws}=-\varepsilon$, and at the equinox declination $\delta_{eq}=0$, during the summer solstice $\delta_{ss}=\varepsilon$.

The measured distance from the center of the ellipse to the center of a cross-shaped sign at the bottom of Varvarinsky slab $m_{meas} \approx 14.8$ cm. In our view, the accuracy of marking on Varvarinsky slab does not exceed ±0.2÷0.3 cm, so is possible to admit that $m_{meas} \approx m_{Tav}$. However, we must bear in mind that changing minor axis *m* on 0.2÷0.3 cm at a fixed semi-major axis $M \approx 20.5$ cm, leads to change the latitude on $\approx 1°$.

Calculated for a plate from Tavriya-1 distance Z, on which necessary to move the gnomon in the winter and summer solstices, respectively: $Z_{ws}$=-6.15 cm, a $Z_{ss}$=6.15 cm.

For 3000 BC, calculated by us the angle of inclination of the ecliptic to the celestial equator is ε = 24024'02 ", and the distances at which it is necessary to move the gnomon in the winter and summer solstice are, respectively, $Z_{ws}$=-6.2 cm и $Z_{ss}$=6.2 cm. They are different from calculated for 1400 BC on 0.05 cm, which is smaller than the measurement error in the ±0.1 cm, which is the minimum scale interval of the measuring ruler. Therefore, it can be argued that the markup of analemmatic sundial for 1400 BC and 3000 BC practically must coincide.

Considering that the minor and major axes of the ellipses of the slab from Tavria-1 and Varvarinsky slab are approximately equal, then the values $Z_{ws}$ and $Z_{ss}$ for both slabs will be identical. Exactly on such a distance the top of the triangle, chipped and darker, than the main surface, area is in the center of Varvarinsky slab. This area is clearly visible in Figure 11.

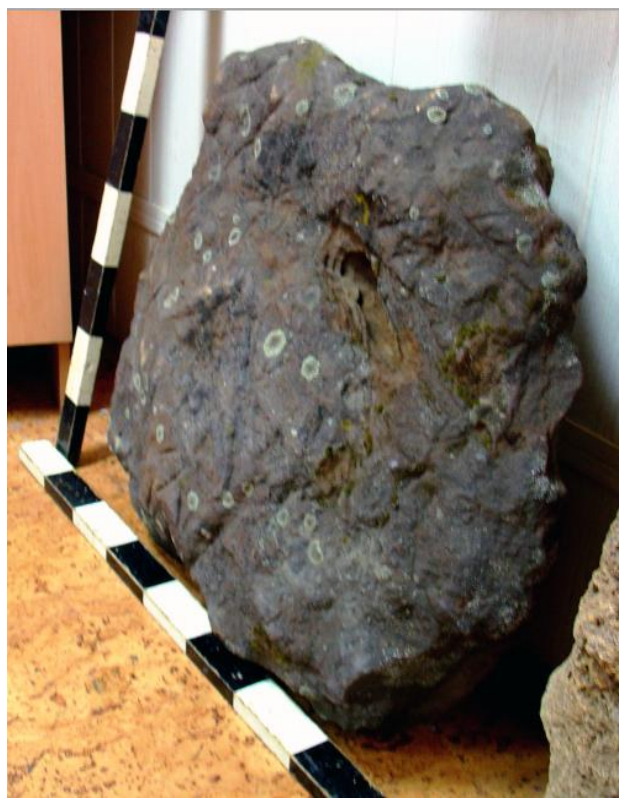

**Figure 11.** Kurgan field Varvarinsky I, kurgan 1 (neighborhood), a stone slab with petroglyphs, side view (Photo by L.N.Vodolazhskaya, 2014). Slab is worth leaning against the wall, without the support of special fasteners (not perpendicular to the floor).

In this area traced several beaten out grooves diverging radially from the top of the triangular area. The angle of its top approximately corresponds to the angular distance between hour lines of analemmatic sundial for the time of sunrise and sunset at the winter solstice.

Hour angles of analemmatic sundial, calculated by the formula 6 for the latitude of Tavriya-1 Lat = 47 ° 17 'N, are shown in Table 1.

**Table 1.** Coordinates of hour marks of analemmatic sundial for latitude *Lat*=47°17' *N*.

*H* - hour angle of the Sun, *H'* - the angle between the noon line and the hour line on the sundial, *t* - time, *x* - coordinate of the hour marks on the axis *X*, *y* - coordinate of the hour marks on the axis *Y*.

|  | *t, hour* | | | | | | | | | | | | |
|---|---|---|---|---|---|---|---|---|---|---|---|---|---|
|  | 6 | 7 | 8 | 9 | 10 | 11 | 12 | 13 | 14 | 15 | 16 | 17 | 18 |
| *H, °* | -90.0 | -75.0 | -60.0 | -45.0 | -30.0 | -15.0 | 0.0 | 15.0 | 30.0 | 45.0 | 60.0 | 75.0 | 90.0 |
| *H', °* | -90.0 | -78.9 | -67.0 | -53.7 | -38.2 | -20.0 | 0.0 | 20.0 | 38.2 | 53.7 | 67.0 | 78.9 | 90.0 |
| *x, cm* | -19.0 | -18.4 | -16.5 | -13.4 | -9.5 | -4.9 | 0.0 | 4.9 | 9.5 | 13.4 | 16.5 | 18.4 | 19.0 |
| *y, cm* | 0.0 | 3.6 | 7.0 | 9.9 | 12.1 | 13.5 | 14.0 | 13.5 | 12.1 | 9.9 | 7.0 | 3.6 | 0.0 |

The exact time of sunrise and sunset in the days of summer and winter solstice were calculated by an astronomical program RedShift 7 Advanced. Results are presented in Table 2.

**Table 2.** Times of sunrise and sunset in the days of solstices for the latitude *Lat*=47°17' *N*.

$t_{rise}$ - the true local solar time at the moment of sunrise, $t_{set}$ - the true local solar time at the moment of sunset, $H'_{rise}$ - the angle between the meridian line and the hour line on analemmatic sundial at the moment of sunrise, $H'_{set}$ - the angle between the meridian line and the hour line at the moment of sunset.

| Phenomenon | $t_{rise}$, hour | $H'_{rise}$, ° | $t_{set}$, hour | $H'_{set}$, ° |
|---|---|---|---|---|
| Summer solstice | 4:03 | -112.4 | 19:56 | 112.1 |
| Winter solstice | 7:48 | -69.5 | 16:13 | 69.7 |

The angle, formed by the shadow of the gnomon from the moment sunrise until sunset during the day of the winter solstice, is highlighted in Figure 12 as the triangular area of gray color.

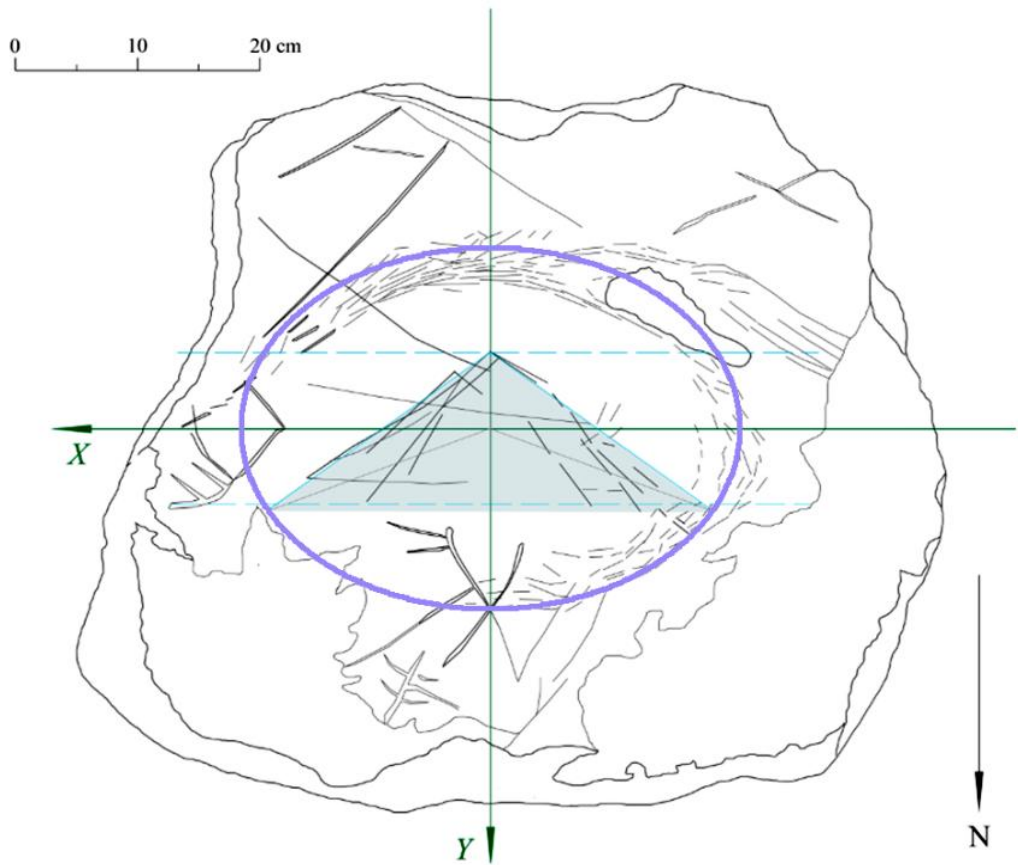

**Figure 12.** Kurgan field Varvarinsky I, kurgan 1 (neighborhood), a stone slab with petroglyphs. The blue ellipse - contour of the "dial" of analemmatic sundial for the latitude Tavriya-1, the apex of the triangular area (highlighted in gray) formed by the shadow of the gnomon from the moment sunrise until sunset during the winter solstice day.

The line, indicating the distance the *Z*, to which it was necessary to move the gnomon at the summer solstice, coincides with the lower offshoot of petroglyph "branch" located below the "rhombus". We assumed that each of the branches of this "branch" may be associated with the same lines *Z* for different latitudes and similar branches in the petroglyph "cross" in the northern part of the slab, respectively, could mark the different values of small semi-axes m.

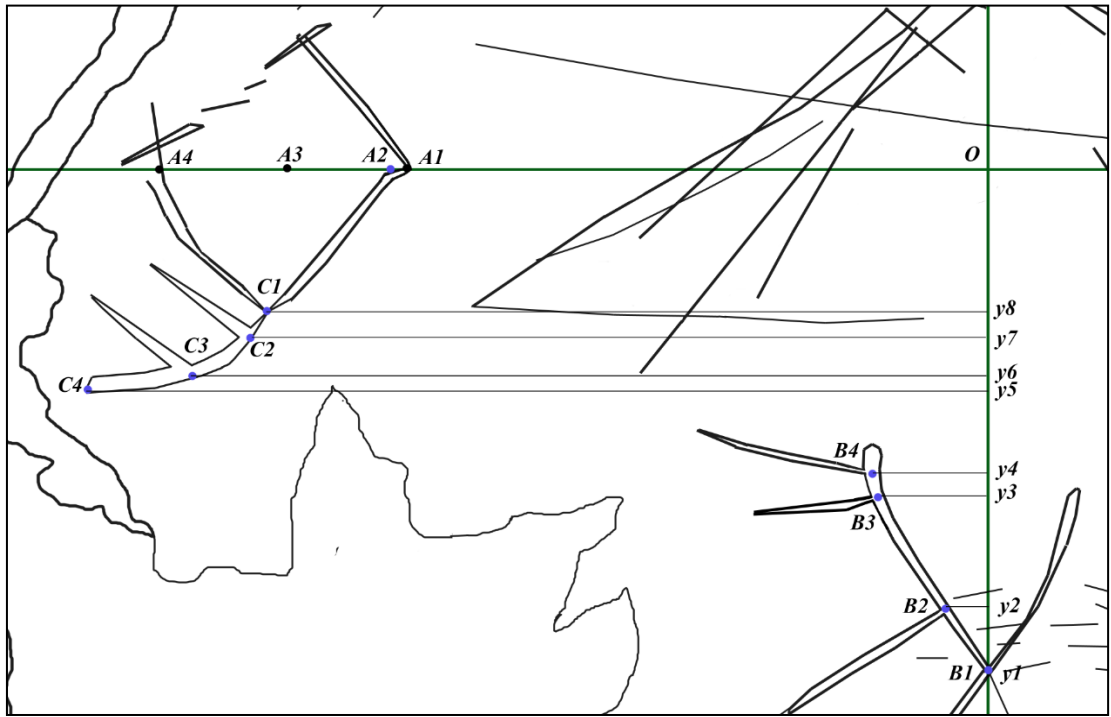

**Figure 13.** Kurgan field Varvarinsky I, kurgan 1 (neighborhood), a stone slab with petroglyphs, fragment of drawing with applied: *A1-A4* points, against which we measured the semi-major axis *M*, *B1-B4* points, which is measured with respect to minor axis *m*, and *C1-C4* points, against which to measure the distance $Z_{ss}$.

As an possible values semi-major axis $M_{meas}$ we were chosen and measured values corresponding to the center and the corners of the "rhombus" in the east of slab - the distance from point *O* to the points *A1, A2, A3, A4* (Fig. 13). The measured values are, respectively: 17 cm, 17.5 cm, 20.5 cm, 24.2 cm Then we have measured the possible minor axes $m_{meas}$ - distances on the *Y* axis from *O* point to points *B1, B2, B3, B4* – values: *y1, y2, y3, y4*. The measured values are, respectively: 14.8 cm, 13.2 cm, 9.7 cm, 9.0 cm.

The calculation of the distance *Z* by the formulas 4 and 5 for the various combinations of the measured values of *M* and *m*, we compared them with the measured $Z_{meas}$ - values y8, y7, y6, y5, which are, respectively: 4.2 cm, 5.0 cm, 6.1 cm, 6.5 cm .

The results of calculations of distances *Z* are shown in Table 3.

**Table 3.** The calculated values of the distance *Z*, on which move the gnomon of analemmatic sundial in the summer solstice, at the different latitudes for given values of the semi-axes *M* and *m*.

| *m, cm* \ *M, cm* | *Z, cm* | | | |
|---|---|---|---|---|
| | 17.0 | 17.5 | 20.5 | 24.2 |
| 14.8 | 3.7 | ***4.1*** | **6.3** | 8.5 |
| 13.2 | 4.7 | ***5.1*** | **6.9** | 9.0 |
| 9.7 | 6.2 | ***6.4*** | 8.0 | 9.7 |
| 9.0 | 6.4 | ***6.6*** | 8.1 | 9.9 |

However, it should be noted that the value of Z is more sensitive to the accuracy of the application petroglyph than *M* and *m*. For mid-latitudes change in the Z on ≈1 mm will change the coordinates of latitude by 1°. Since the precision of all petroglyphs of slab likely was the same, then the branch lines will correspond specify Z with an accuracy ±2÷3 mm and give an accuracy ±2÷3°. This is a rather large error, so the "branch" lines can be viewed only as auxiliary lines for the approximate determination of the distance Z on particular latitude.

Calculations of latitude were made by us on the formula 9 obtained from the formula 1:

$$\varphi = arcsin(m/M), \qquad (9)$$

where *φ* - latitude of location, *M* - semi-major axis of the ellipse, *m* - minor axis of the ellipse.

The results of these calculations are presented in Table 4.

**Table 4.** The calculated values of the latitude *φ*, for which analemmatic sundial with given values of semi-major axis *M* and the minor axis *m* are intended.

| M, cm / m, cm | φ, ° | | | |
|---|---|---|---|---|
|  | 17.0 | 17.5 | 20.5 | 24.2 |
| 14.8 | 60.5 | **57.7** | **46.2** | 37.7 |
| 13.2 | 50.9 | **49.0** | **40.1** | 33.1 |
| 9.7 | 34.8 | **33.7** | 28.2 | 23.8 |
| 9.0 | 32.0 | **30.9** | 26.0 | 21.8 |

**The discussion of the results**

Analyzing the results, we concluded, that in case the value of semi-major axis of the ellipse *M*=24.2 cm calculated distance Z (see. Tab. 3) is significantly higher than the distance at which there are offshoots on the "branch", so the use of semi-major axis of this length was unlikely.

For the semi-major axis *M*=20.5 cm in two cases calculated values Z almost coincide with the measured (*C3* and *C4* point) (see. Tab. 3). They correspond to small semiaxes 14.8 cm and 13.2 cm, for which latitude are ≈46° N (approx the latitude of Tavria-1 or the Northern Coast of the Black Sea) and ≈40° N (approx the latitude of the Southern Black Sea Coast) (see. Tab. 4).

If the wells of the slab of Tavria-1 put on drawing of Varvarinsky slab, then the ellipse of analemmatic sundial, corresponding latitude of Tavria-1, on the north side will be very accurately touch the outer edge of the wells (Fig. 14). However, the main the wells: Eastern (18 o'clock), Western (6 o'clock) and South (24 o'clock), it crosses almost in the center. Perhaps such an arrangement wells was connected with the attempt to create a more universal markup of a sundial. Center of the 12 o'clock well located approximately at the level of the point B2 of the first offshoot of petroglyph "skew cross" (Fig. 13). This point is related to the minor axis m≈13.2 cm, roughly corresponding to the latitude of the Southern Black Sea Coast - ≈40° N. Ellipse of the "dial" of analemmatic sundial with *M*=20.5 cm and *m*=13.2 cm crosses almost the middle of most of wells of operating range from 6 to 18 o'clock. Those, to measure the time in this range for the Southern Coast of the Black Sea had to be oriented mainly on the centers of wells. Thus, the sundial on the slab from Tavriya-1 made it possible to determine the time, as on the North as well on the South Coast of the Black Sea.

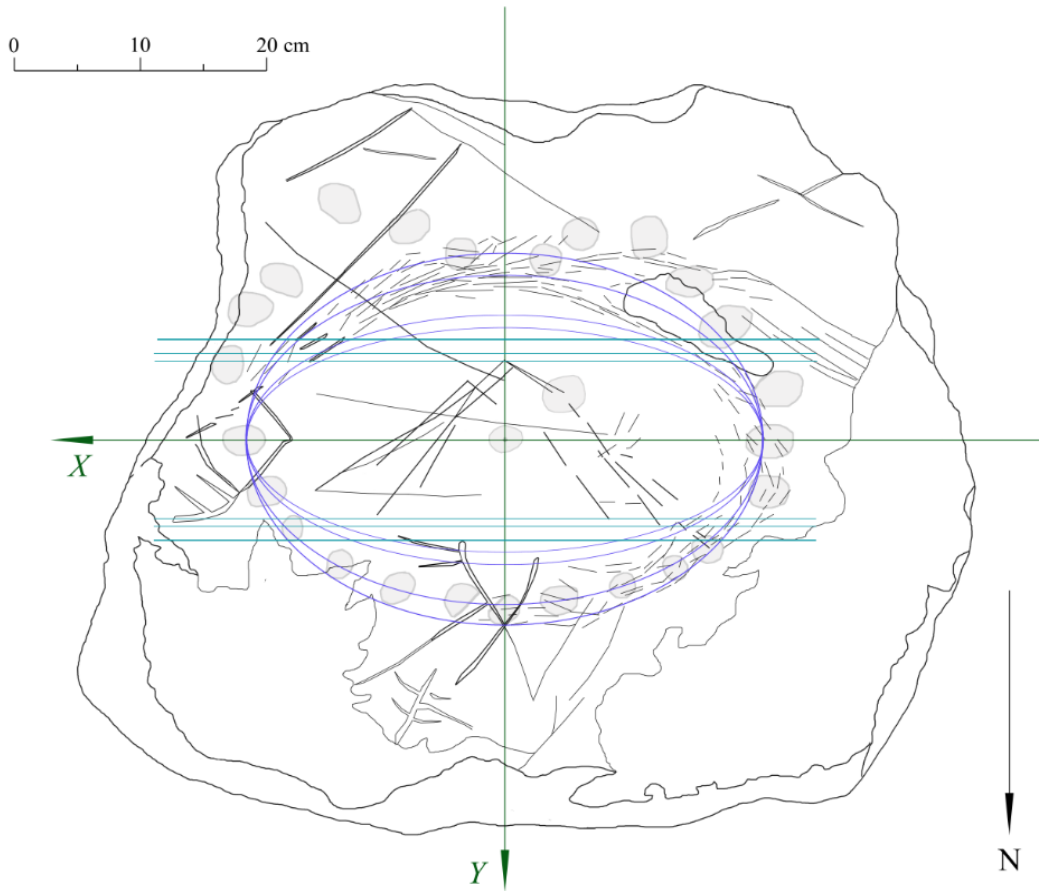

**Figure 14.** Kurgan field Varvarinsky I, kurgan 1 (neighborhood), a stone slab with petroglyphs, ellipse of analemmatic sundial "dial" with semi-major axis *M*=20.5 cm and variety small semi-axes *m* highlighted in blue, the green line mark the corresponding distances $Z_{ws}$ and $Z_{ss}$.

Accordingly, Varvarinsky slab could serve as a model for markup of analemmatic sundial "dial" ellipse (with large semi-axis of the ellipse *M*≈20.5 cm) for latitudes of all coast of the Black Sea.

Fully, although with small errors, were involved all offshoots of "branch" for the semi-major axis *M*=17.5 cm (see tab. 3), representing the segment from the center of the ellipse O to the closest to it a corner of "rhombus", however, not the point *A1*, but point *A2*. Point *A1* appeared due to not very precise application of one of the lines "rhombus" likely (Fig. 13).

For the semi-major axis *M*=17.5 cm proved to involved all points (Fig. 15), the corresponding values of minor semiaxis: 14.8 cm, 13.2 cm, 9.7 cm, 9 cm, latitudes, approximately: 57.7º N, 49.0º N, 33.7º N and 30.9º N ( see. tab. 4).

Latitude, 57.7ºN, close to the latitude of 'belye nochi' ('white nights'), equal ≈60.5ºN, where in summer there are periods when civil twilight (sun descends less than 6º below the horizon) continued throughout the night and, thus the separation of day on the light and dark parts violated.

Latitude ≈49.0º *N* near latitude 48.5º *N*, south of which, even near the summer solstice, is observed an astronomical night (Sun descends by more than 18º below the horizon). From this latitude, and to the south, in the summer solstice night in the sky is already possible to observe even weak stars and the Milky Way. North of this is not possible. Latitude ≈49.0º *N*, also close to the latitude of the place of Varvarinsky slab detection equal ≈49.5º *N*.

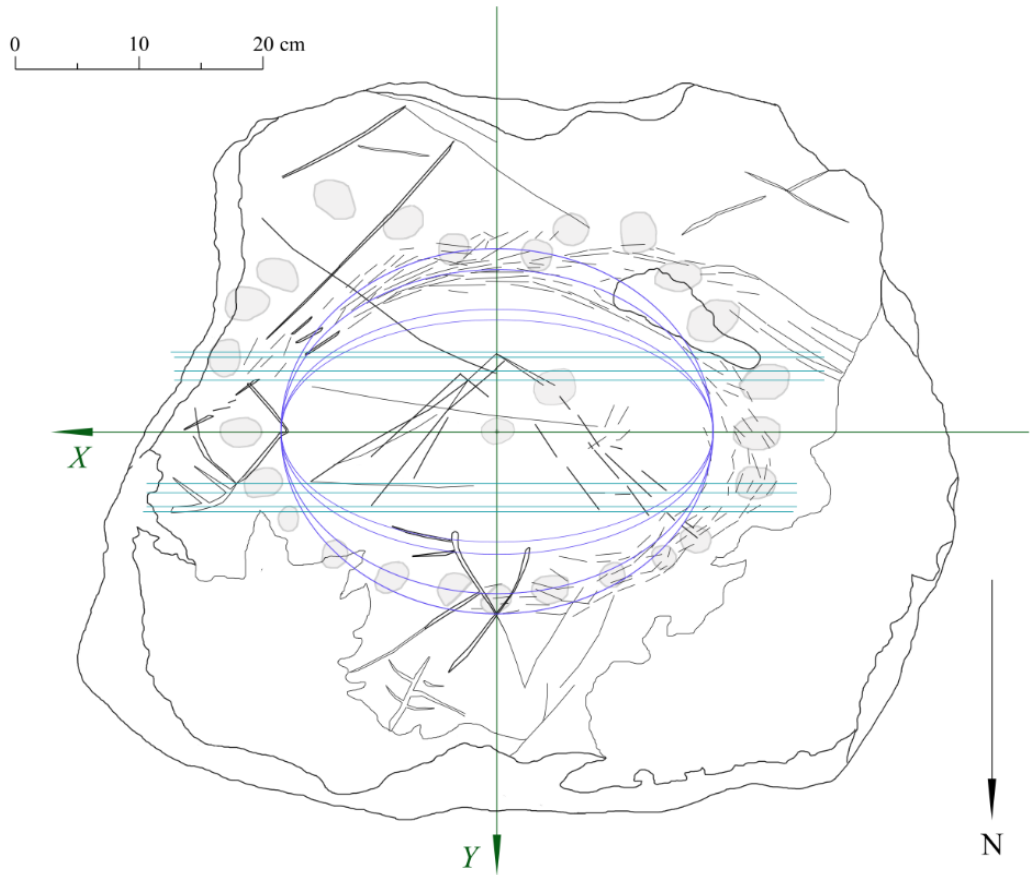

**Figure 15.** Kurgan field Varvarinsky I, kurgan 1 (neighborhood), a stone slab with petroglyphs, ellipse of analemmatic sundial "dial" with semi-major axis *M*=17.5 cm and variety small semi-axes *m* highlighted in blue, the green line mark the corresponding distances $Z_{ws}$ and $Z_{ss}$.

Calculated hour markers for these northern latitudes (57.7º *N* and 49.0º *N*) fit well into the area of the elliptical groove of Varvarinsky slab, so Varvarinsky slab could be used as a model for making / markup ellipse of analemmatic sundial "dial" with *M*=17.5 cm in the range of northern latitudes (from the border surveillance astronomical night in the days of the summer solstice to the border "white nights" with a complete absence of the possibility of observation of the starry sky near the summer solstice).

Latitude 33.7º *N* is much to the south. Close to this latitude are located islands Crete (≈35÷35.5º *N*) and Cyprus (≈34.5÷35.5º *N*). Latitude 30.9º *N* falls in latitudes range from the 30.0º *N* to 31.5º *N*, in which the Nile Delta is located. These latitudes correspond to the marks on the petroglyphs of "rhombus" and "oblique cross", but ellipses missing on the slab for these latitudes. Marks on the petroglyphs "rhombus" and "oblique cross" correspond to these latitudes, but there are no ellipses on the slab for these latitudes. This suggests that, most likely, Varvarinsky slab is not used to these latitudes. Although, if necessary, it was possible get the basic dimensions of the ellipse of analemmatic sundial "dial" (and distances Z) for such southern latitudes with its help.

Thus the study of Varvarinsky slab and analysis of petroglyphs, applied onto it, allowed a conclusion that Varvarinsky slab is a prototype of Srubna slab from burial of kurgan field Tavriya-1. Petroglyph in the form of elliptical groove on the Varvarinsky slab in its parameters corresponds to the elliptic location of wells - hour marks of analemmatic sundial on the slab from the Tavriya-1. However, there are no wells on the Varvarinsky slab, which could serve as hour marks, so it can not be regarded as a sundial. It is a prototype of analemmatic sundial, markup which is stamped on the slab from the Tavria-1 and precedes the slabs with petroglyphs from the Srubna burials - analemmatic sundials discovered in the Rostov and Donetsk Oblast.

Is most likely, that Varvarinskaya slab with petroglyphs used for developing the method of a markup of analemmatic sundials. The discovery of the prototype of a sundial in approximately the same area where there were found all the currently known Srubna analemmatic sundials, testifies in favor of local origin, as the very idea of analemmatic sundials, well as their markup technology.

**Acknowledgements**

The authors express their sincere gratitude to A.V. Fayfert, A.N. Usachuk, V.A. Larenok, V.F. Chesnok, employees of archaeological department GAUK RO "Donskoe nasledie" and employees of the Archaeological Museum-Reserve "Tanais" for the support of research.

**List of Abbreviations**

GAUK RO – Gosudarstvennoe avtonomnoe uchrezhdenie kul'tury Rostovskoy oblasti [State independent establishment of Culture of Rostov Oblast];
TAE – Taganrogskaya arkheologicheskaya ekspeditsiya [Taganrog archaeological expedition];
TGLIAMZ – Taganrogskiy gosudarstvennyy literaturnyy i istoriko-arkhitekturnyy muzey-zapovednik [Taganrog State Literary and Historical-Architectural Museum-Reserve].